\documentclass{ws-procs9x6}
\begin{document}
%self-made expressions
\newcommand{\dzero}     {D\O~}
\newcommand{\dzerons}     {D\O}
\newcommand{\ttbar}     {$t\overline{t}$~}
\newcommand{\ppbar}     {$p\overline{p}$~}
\newcommand{\qqbar}     {$q\overline{q}$~}
\newcommand{\MET}       {\mbox{$\not\!\!E_T$}~}
\newcommand{\METns}     {\mbox{$\not\!\!E_T$}}
\newcommand{\pt}        {$p_T$~}
%%%%%%%%%%%%%%%%%%%%%%%%%%%%%%%%%%%%%%%%%%%%%%%%%%%%%%%%%%%%%%%%%%%%%%%%%%%%%%%%%%%%%%%%%%%%%%%%
\title{D\O~Top Physics}
\author{Marc-Andr\'e Pleier}
\address{University of Rochester\\
E-mail: pleier@fnal.gov\\
for the \dzero collaboration}
\maketitle
%%%%%%%%%%%%%%%%%%%%%%%%%%%%%%%%%%%%%%%%%%%%%%%%%%%%%%%%%%%%%%%%%%%%%%%%%%%%%%%%%%%%%%%%%%%%%%%%
\abstracts{The Tevatron proton--antiproton collider at Fermilab operates at a 
centre of mass energy of 1.96 TeV and is currently the only source 
for the production of top quarks. 
Recent \dzero results on
the top quark's production cross section and its properties
such as mass, helicity of the {\it W} in its decay and branching fraction 
B($t \rightarrow Wb$) are presented,
and probe the validity of the Standard Model (SM).}
%%%%%%%%%%%%%%%%%%%%%%%%%%%%%%%%%%%%%%%%%%%%%%%%%%%%%%%%%%%%%%%%%%%%%%%%%%%%%%%%%%%%%%%%%%%%%%%%
\section{Introduction}
The top quark, which completes the quark sector of the SM, was
discovered in 1995 at the Tevatron by the CDF and \dzero
Collaborations~[\refcite{top_discovery}].
Being the heaviest of all quarks with a mass of 178.0 $\pm$ 
4.3 GeV/c$^2$~[\refcite{topmass}], it couples most strongly
to the Higgs boson, and its lifetime of $\approx 4\cdot 10^{-25}$~s 
means that it is the only quark that decays before it can hadronise, preserving spin information,
and providing a way to study the decay of an essentially free quark. 
Measuring the production cross section of the top quark and its different 
properties such as  mass, {\it W} helicity in its decay, branching fraction 
B($t \rightarrow Wb$), {\it etc.}, and comparing with
predictions of the SM is a very powerful tool for searching for physics
beyond the SM. 
%%%%%%%%%%%%%%%%%%%%%%%%%%%%%%%%%%%%%%%%%%%%%%%%%%%%%%%%%%%%%%%%%%%%%%%%%%%%%%%%%%%%%%%%%%%%%%%%
\section{Top quark production at the Tevatron}
In \ppbar collisions at a centre of mass energy $\sqrt{s}=1.96$ TeV, 
top quarks are produced predominantly in pairs: \ppbar $\rightarrow$ \ttbar + X via 
the strong interaction (85\% \qqbar annihilation and 15\% gluon-gluon fusion).
At next-to-next-to-leading order, the corresponding SM cross section is
6.77 $\pm$ 0.42 pb~[\refcite{ttbar_xsec}].
According to the SM, the top quark decays predominantly into
{\it W} bosons and {\it b} quarks. Hence, there are three event classes to be 
observed resulting from \ttbar decay, which depend on the decay mode
of the {\it W} bosons: \mbox{({\it i}) a dilepton} final state where both {\it W} bosons decay leptonically,
resulting in two isolated high-\pt leptons, missing transverse energy \MET corresponding to the two
neutrinos and two jets, ({\it ii}) a lepton+jets final state where one {\it W} boson
decays leptonically, the other one hadronically, resulting in one isolated 
high-\pt lepton, \MET and four jets, and ({\it iii}) an all-jets final state where both
{\it W} bosons decay to $\overline{q}q'$ pairs producing six jets. In all final states, two
of the jets are {\it b}-jets, and additional jets can arise from ISR/FSR. The all-jets final state represents the biggest 
branching fraction of \ttbar events ($\approx$46\%), but it is also difficult to separate
from a big multijet background. The dilepton final state without $\tau$ leptons constitutes $\approx$5\%
of the \ttbar events and gives the cleanest signal, but also suffers from 
low statistics. The lepton+jets events in the $e$+jets or $\mu$+jets channels yield $\approx$29\% of the branching
fraction and provide the best compromise between sample purity and statistics.

In addition to \ttbar pair production, top quarks can also be produced singly
via the electroweak interaction through a {\it Wtb} vertex.
A measurement of single top quark production consequently provides direct access to the CKM
matrix element $V_{tb}$. 
Depending on the virtuality
(squared four-momentum) of the participating {\it W} boson ($Q^2_W$), there are two contributions:
the s-channel $q'\bar{q}\rightarrow t\bar{b}$ ($Q^2_W > 0$), with a predicted cross section of
0.88 $\pm$ 0.05 pb~[\refcite{singletop_s_xsec}], and the t-channel  
$q'g \rightarrow tq\bar{b}$ ($Q^2_W < 0$), with a predicted cross section of 1.98 $^{+0.28}_{-0.22}$ pb
~[\refcite{singletop_t_xsec}].
The contribution from single top production in both s and t channels of $bg\rightarrow tW$
can be neglected at Tevatron energies [\refcite{singletop_tW}].

We present only
results from leptonic channels containing a muon or an electron, and
$W\rightarrow\tau$ decays are therefore included in a partial way, depending on the $\tau$ decay mode.
%%%%%%%%%%%%%%%%%%%%%%%%%%%%%%%%%%%%%%%%%%%%%%%%%%%%%%%%%%%%%%%%%%%%%%%%%%%%%%%%%%%%%%%%%%%%%%%%
\section{Measurement of the $\rm t\bar{t}$ production cross section}
The \ttbar pair production cross section has been measured by \dzero in 
several decay modes, using either purely topological and kinematic
event properties to separate the \ttbar signal from background,
or by adding identification of {\it b}-jets based mainly on the long lifetime of 
B hadrons. Several algorithms are deployed for {\it b}-jet identification, {\it e.g.},
searching for muons in jets resulting from semileptonic B decays (soft-$\mu$ tag) or
using reconstructed secondary vertices (SVT),
or the significance of impact parameters of tracks within jets relative to
primary vertices (CSIP). The probability to tag at least one jet in a \ttbar lepton+jets
event by its lifetime is $\approx$60\%, whereas the main background from {\it W}+jets production 
is tagged in only $\approx$4\% of the cases, resulting in an improved signal to
background ratio in tagged analyses. An advantage of topological analyses
is that they do not depend on the assumption of 100\% branching of B($t \rightarrow Wb$),
and are therefore less model-dependent than tagging analyses. Analysing different decay channels 
helps to improve statistics of top events, 
and studies of properties, as well as probing of physics beyond
the SM that could result in enhancement/depletion in some particular channel.
Figure~\ref{plot:ttbar_xsec} shows an overview of all cross section measurements performed thus far
at \dzerons. We find that all measurements are in good agreement with the SM
and with each other.
%%%%%%%%%%%%%%%%%%%%%%%%%%%%%%%%%%%%%%%%%%%%%%%%%%%%%%%%%%%%%%%%%%%%%%%%%%%%%%%%%%%%%%%%%%%%%%%%
\section{Measurement of the top quark mass}
\begin{figure}[ht]
\centerline{\epsfxsize=2.5in\epsfbox{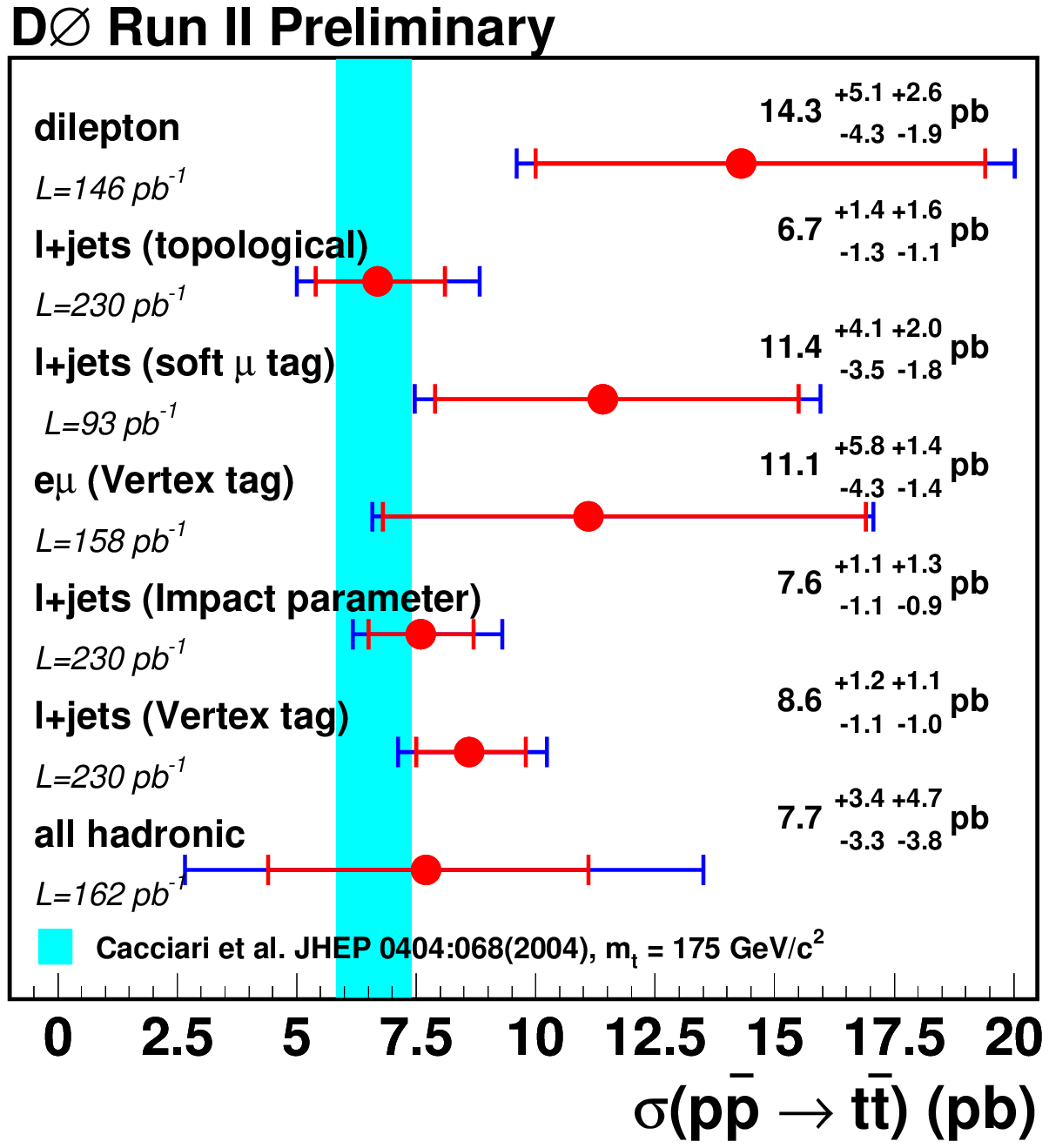}\epsfxsize=2.5in\epsfbox{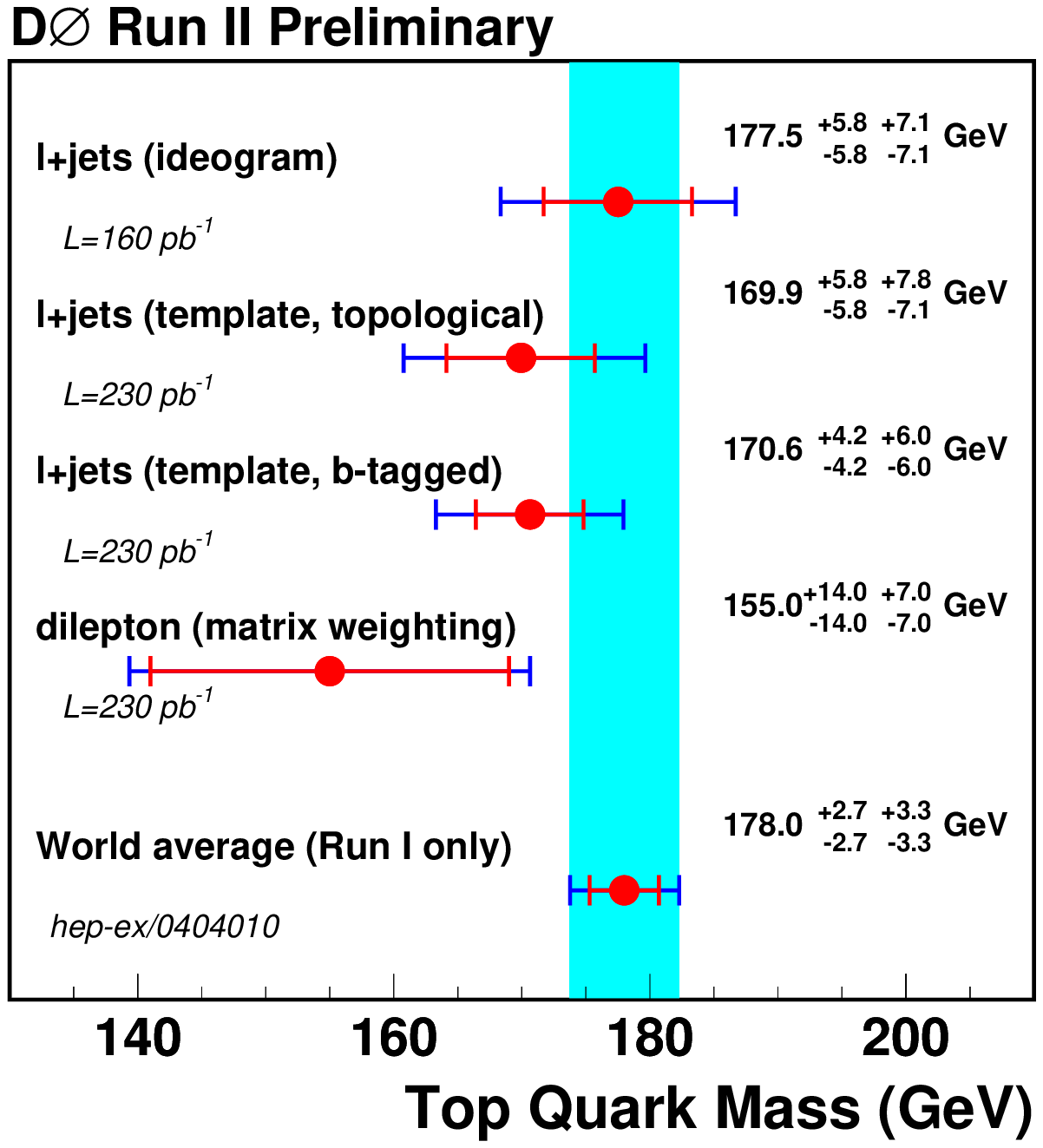}}   
\caption{Left: \ttbar pair production cross section as measured by \dzero in Run II, and the SM
prediction.\label{plot:ttbar_xsec}
Right: Top quark mass measurements from \dzero in Run II compared
to the world average based on Run I measurements. \label{plot:ttbar_mass}}
\end{figure}

The top quark mass is a fundamental parameter that is not predicted by the SM, but 
can be used together with the {\it W} mass to constrain the mass of 
the Higgs boson via radiative corrections.
\dzero has measured the top quark mass in the lepton+jets and dilepton channels
using different techniques:
In the lepton+jets channel, a kinematically-constrained fit is used to
extract the top quark mass from preselected events, using either template mass spectra
for signal and background (template method) or an analytical likelihood
for calculating the probability for any event to be signal or background (ideogram method).
In the dilepton channel, of the 6-particle final state, only four objects 
are detected together with \METns, which provides an underconstrained problem. As proposed
by Dalitz, Goldstein~[\refcite{dilepton_mass_dalitz}] and Kondo~[\refcite{dilepton_mass_kondo}], 
a hypothesised value of the top quark mass can be used to solve for the \ttbar momenta.
The solutions yield a weight distribution for each preselected event as a function
of the top quark mass. Using its peak as an estimator of the mass for each event, and comparing the resulting
distribution from all preselected events to signal and background templates, 
provides an estimate of the top quark mass.
All the measurements are shown in Fig.~\ref{plot:ttbar_mass}, and all 
are in agreement with the current world average~[\refcite{topmass}].
The final goal is to measure the top quark mass with 1\% precision in Run~II.
%%%%%%%%%%%%%%%%%%%%%%%%%%%%%%%%%%%%%%%%%%%%%%%%%%%%%%%%%%%%%%%%%%%%%%%%%%%%%%%%%%%%%%%%%%%%%%%%
\section{Measurement of the {\it W} helicity in $\rm t\bar{t}$ decays}
Top quark decay in the V$-$A charged current weak interaction proceeds only via a left-handed ($f^{-}$=
30\%) and a longitudinal ($f^{0}$=70\%) fraction of {\it W} helicities, which is reflected
in the angular distribution of the charged lepton relative to the line of flight of the top quark
in the {\it W} rest frame in lepton+jets final states.
Using data corresponding to integrated luminosities of 169 pb$^{-1}$ 
($e$+jets) and 158 pb$^{-1}$ ($\mu$+jets), and comparing the above mentioned
angular distribution in data to templates, where we set $f^0$ to its SM
value, and vary the right-handed fraction $f^+$, and correspondingly 
$f^-$, between 30\% and 0\%,
we obtain an upper limit on $f^+$ from a binned likelihood fit of:
\begin{eqnarray}
  f^+<0.24~{\rm (90\%~CL),}\nonumber
\end{eqnarray}
in agreement with expectation from the SM.
%%%%%%%%%%%%%%%%%%%%%%%%%%%%%%%%%%%%%%%%%%%%%%%%%%%%%%%%%%%%%%%%%%%%%%%%%%%%%%%%%%%%%%%%%%%%%%%%
\section{Measurement of B($t \rightarrow Wb$) / B($t \rightarrow Wq$)}
The ratio of branching fractions R =  B($t \rightarrow Wb$) / $\Sigma_{q=d,s,b}$ 
B($t \rightarrow Wq$) is constrained within the SM to $0.9980 < R < 0.9984$  
at 90\% CL~[\refcite{R_range}], assuming three fermion generations, unitarity of 
the CKM matrix and neglect of non-{\it W} decays of the top quark.
\dzero has measured R in the lepton+jets channel using data corresponding
to integrated luminosities of 169 pb$^{-1}$ ($e$+jets) and 158 pb$^{-1}$ ($\mu$+jets),
by comparing the number of single to double {\it b}-tagged events, using one {\it b}-jet to 
identify the \ttbar event and the second one for the measurement of the relative 
fraction of $t \rightarrow Wb$. 
For example, applying the SVT algorithm for {\it b}-tagging,
we obtain the following result for R from a simultaneous fit of R and the
\ttbar cross section:
\begin{eqnarray}
  R = 0.70^{+0.27}_{-0.24}\:{\rm (stat)}\:^{+0.11}_{-0.10}\:{\rm(syst).}\nonumber
\end{eqnarray}
This agrees with the SM expectation.
%%%%%%%%%%%%%%%%%%%%%%%%%%%%%%%%%%%%%%%%%%%%%%%%%%%%%%%%%%%%%%%%%%%%%%%%%%%%%%%%%%%%%%%%%%%%%%%%
\section{Search for single top quark production}
Thus far, the search for single top quark production
has been performed only for the cases where the {\it W} boson decays into
an electron or muon and neutrino, resulting in final states
with one isolated high-\pt lepton, \MET and two or three jets.
Using data for an integrated luminosity of 230 pb$^{-1}$,
with selections optimised for
leptonic {\it W} decays, maximal acceptance of signal and good 
modelling of the remaining backgrounds, a set of common
discriminating variables is used to separate signal
from the background. One analysis is based on classical cutoff criteria, optimised via
a random grid search, and another on a multivariate (neural nets/decision
trees) analysis.
The neural network analysis gives the best limits:
\begin{eqnarray}
 \sigma_s < 6.4\:{\rm pb\: (95\% \:CL)}\nonumber,~~~~
 \sigma_t < 5.0\:{\rm pb\: (95\% \:CL).}\nonumber
\end{eqnarray}
These limits represent a factor of two improvement
relative to all other measurements, and are in agreement with expectations from the SM. 
%%%%%%%%%%%%%%%%%%%%%%%%%%%%%%%%%%%%%%%%%%%%%%%%%%%%%%%%%%%%%%%%%%%%%%%%%%%%%%%%%%%%%%%%%%%%%%%%
\section{Summary}
A wealth of top analyses is being pursued at \dzerons, continuing to probe
the validity of the SM. So far, all measurements are in agreement with the SM.
More detailed descriptions of the analyses
can be found online~[\refcite{top_group_web}].
Continuously improving the analysis methods, and using the increasing 
integrated luminosity from a smoothly running Tevatron, expected to deliver more than 4~fb$^{-1}$ 
by the end of Run~II, we are moving towards precision measurements and
hopefully discoveries within and outside the SM.
%%%%%%%%%%%%%%%%%%%%%%%%%%%%%%%%%%%%%%%%%%%%%%%%%%%%%%%%%%%%%%%%%%%%%%%%%%%%%%%%%%%%%%%%%%%%%%%%

\end{document}